\documentclass[twocolumn,showpacs,preprintnumbers,amsmath,amssymb,epsfig,widetext]{revtex4}

\usepackage{graphicx}% Include figure files
\usepackage{dcolumn}% Align table columns on decimal point
\usepackage{bm}% bold math
\usepackage{epsfig}

\def\fun#1#2{\lower3.6pt\vbox{\baselineskip0pt\lineskip.9pt
  \ialign{$\mathsurround=0pt#1\hfil##\hfil$\crcr#2\crcr\sim\crcr}}}

\input epsf

\def\be{\begin{equation}}
\def\ee{\end{equation}}
\def\bi{\begin{itemize}}
\def\ei{\end{itemize}}
\def\ba{\begin{eqnarray}}
\def\ea{\end{eqnarray}}

\def\nn{\nonumber}

\newcommand{\ie}{{\it i.e. }}

\def\hatn{{\bf \hat n}}

\def\veck{{\bf k}}

\def\ie{{\it i.e.~}}

\long\def\comment#1{}

\begin{document}

\preprint{}

%============================================================================
\title{A step towards testing general relativity using weak gravitational lensing and redshift surveys}
%============================================================================

\author{Yong-Seon Song$^{1}$ and Olivier Dor\'e$^{2}$}
\email{Yong-seon.song@port.ac.uk,olivier@cita.utoronto.ca}
\affiliation{$^{1}$Institute of Cosmological $\&$ Gravitation, University of Portsmouth, Portsmouth, PO1 3FX, UK }
\affiliation{$^{2}$Canadian Institute for Theoretical Astrophysics, 60 St. George St, University of Toronto, Toronto, ON, Canada M5S3H8}

\date{\today}

\begin{abstract}
Using the linear theory of perturbations in General Relativity, we
express a set of consistency relations that can be observationally
tested with current and future large scale structure
surveys. We then outline a stringent model-independent program to test gravity on
cosmological scales. We illustrate the feasibility of such a program
by jointly using several observables like peculiar velocities, galaxy
clustering and weak gravitational lensing. After addressing possible observational or astrophysical
caveats like galaxy bias and redshift uncertainties, we forecast in
particular how well one can predict the lensing signal from a cosmic
shear survey using an over-lapping galaxy survey. We finally discuss the
specific physics probed this way and illustrate how $f(R)$ gravity
models would fail such a test.
\end{abstract}

%\pacs{draft}

%\keywords{CMB-inflation}

\maketitle

\section{Introduction}
%=====================

To understand the origin of the accelerating expansion of the universe
is one of the salient question in contemporary cosmology
\cite{Riess:2004nr,Astier:2005qq,Eisenstein:2005su,Cole:2005sx,Tegmark:2006az,Komatsu:2008hk,Kilbinger:2008gk}. It
is commonly attributed to the existence of some extra unknown physics loosely labelled Dark
Energy (DE) \cite{Peebles:2002gy}. Alternatively, it might cast some doubts on the
theoretical foundation of  current cosmology, that is the theory of General Relativity (GR). From a
theorist point of view, as most efforts to construct self-consistent
DE models within GR seem unsuccessful, there is no a priori reason to
rule out a modification to gravity on cosmological scales as an
explanation \cite{Durrer:2008in}. From an observer point of view, whereas GR passes direct tests probing the Solar System scales ($10^{11}$m) down to
the laboratory scales ($10^{-3}$m) \cite{adelberger03}, testing
gravity on cosmological scales is challenging and overly model dependant so far. In this work, we propose to remedy
these issues by defining a set of simple model independent tests of GR on 
cosmological scales, although more complicated models of GR (clumping
dark energy or interacting dark energy models) are not considered in
this paper (the complete set of test including both exotic dark energy
models is being prepared as the second step of testing gravity). 
The tests we devise rely on simple and testable
consistency relations based on linear GR and make use of the 
appropriate combinations of various observables of large scale
structures. Were any of these relations proved to be violated, it
would be a clear sign of a break-down of GR and it would highly
motivate theoretical work on alternatives to GR. Reversely, before any
such conclusion can be reached, it requires an absolute demonstration
that observational systematic effects are well under controlled. A
firm believer in GR might look at these states as simple test of
systematic effects. Note that throughout this work, we will assume the
validity of standard cosmology, that is to say the validity of the Copernican
principle \cite{Quercellini:2008ty}. If this was not the case, then the DE problem could be
solve without new physics and more specific tests would have to be
designed \cite{Goodman:1995dt,Caldwell:2007yu,Uzan:2008qp}.

Several alternative approaches to test GR have been proposed so
far, some parametric, some non-parametric. The parametric approaches
can be classified into several categories according to their goal: i)
to separate the geometrical and growth signatures in the $w$ $\&$
$w_a$ plane~\cite{Ishak:2005zs,Song:2005gm,Wang:2007fsa}, ii) to
define the probed $w(z)$ range in a model independant manner \cite{Knox:2006fh,Mortonson:2008qy,SongDore08}, iii) to look for
an anomalous linear growth rate using the $\gamma$
parameter~\cite{Linder:2005in,Linder:2007hg,Huterer:2006mva,Polarski:2007rr,DiPorto:2007ym,Mortonson:2008qy,Thomas:2008tp,Dent:2008ia}, iv) to parametrize the metric
perturbations~\cite{uzan01,Song:2006sa,Bertschinger:2006aw,Koivisto:2005mm,Amendola:2007rr,Uzan:2006mf,Hu:2007pj,Hu:2008zd,Zhao:2008bn,Bertschinger:2008zb,Daniel:2008et,Caldwell:2007cw,Dore:2007jh}. Alternatively, non-parametric approaches were developed by~\cite{Jain:2007yk,Song:2008vm,Zhang:2007nk,Zhang:2008ba}. 

In this paper, we advocate the use of a non-parametric approach -- based
on a set of GR based consistency relations -- to test gravity on cosmological
scales. We will use explicit alternative to GR for illustrative
purpose only. To do so, we jointly use various observables, \ie
velocities, galaxy clustering and weak-gravitational lensing~\cite{Jain:2007yk}. More
precisely, by using a combination of the two first, we will be able to
compare it to the later. In this work, extending the work of
\cite{Song:2008vm} we focus closely on potential observational
biases. We propose in particular various ways to deal with the galaxy
bias, redshift space distortion and spectroscopic and photometric redshifts.  

We first begin by introducing in Sec.~\ref{sec:conv} a general
framework to test GR on cosmological scales.  We then detail how to
build faithful probes of matter perturbations in
Sec.~\ref{sec:tracing_matter} before detailing one test of GR in
Sec.\ref{sec:test}. We conclude and discuss our findings in
Sec.~\ref{sec:disc}. 

\section{Gravitational consistency tests on cosmological scales}
%===============================================================
\label{sec:conv}

In a metric theory of gravity, large scale structures
correspond to metric and matter-energy momentum perturbations. In the Newtonian gauge, metric perturbations are described by
\be
ds^2=-(1+2\Psi)dt^2+(1+2\Phi)a^2\delta_{ij}dx^idx^j\,,
\ee
where $\Phi$ and $\Psi$ denote curvature perturbations and the Newtonian
force respectively. 
%In the absence of anisotropic stress,
%\be
%\Phi=-\Psi\ .
%\ee
%Both a departure from general relativity (GR) -- that we should refer to as modified gravity-- or a clustering of dark energy in GR
%could create some anisotropic stress and breaks the equivalence
%between $\Phi$ and $-\Psi$.
In this work (except in section Sec.~\ref{sec:cc_with_vel}) we will aim at constructing GR consistency
relations and thus we will work within GR. We will restrict ourselves
to a linear theory of perturbations on sub-horizon scales to get simpler and easier to test relations. We will
also consider matter density fluctuations to be dominating the growth
of perturbations and thus neglect DE clustering.

The first simple relation that can be tested within GR is based on the
continuity equation: 
\be\label{eq:en_con}
\frac{d\delta_m}{dt}=-\frac{\theta_m}{a}
\ee
where $\theta_m$ is given by $\theta_m=\vec\nabla\cdot \vec v_m$.
Both sides of this equation can be probed observationally
independently and we shall call the observational test of this
relation the {\it energy-momentum consistency test}. As will be seen
below, whereas peculiar velocities can be traced directly, probing matter
density fluctuations requires to deal with several observational
artifacts. In this paper, we will assume this relation to be satisfied
and we will use it for our observational determination of the galaxy
bias. Note however that it could be violated due to dark sector interactions.

The second consistency relation stems from the lack of anisotropic stress and
relates metric perturbations. Since in this paper, we neglect DE
clustering, only the matter component clusters and the no anisotropic
stress approximation is valid, 
\be\label{eq:me_con}
\Phi+\Psi=0\,,
\ee
which reduces the degrees of freedom of metric
perturbations. While the Newtonian force $\Psi$ can be reconstructed from
the evolution of peculiar velocities, the curvature perturbation $\Phi$ is given
by matter fluctuations. Thus both observables could be compared to
determine the presence of non-trivial anisotropic stress which is
predicted in most modified gravity models and dark energy clumping
model. We call this test the {\it metric consistency test}. 

The other tests relate metric perturbations to matter-energy
fluctuations, dynamically or non-dynamically. Newtonian force sources
the dynamics of matter fluctuations. If the time evolution of peculiar
velocity can be reconstructed, then so can $\Psi$ through the Euler equation,
\be\label{eq:dyn_con}
\frac{d\theta_m}{dt}=-H\theta_m + \frac{k^2\Psi}{a}\,,
\ee
which we label as the {\it dynamical constraint test}. Because the degrees
of freedom available in metric perturbations are reduced by the
no-anisotropy condition in GR, we shall not use this constraint in this paper. 

Finally, the relation between curvature perturbations and matter
fluctuations yield the {\it non-dynamical constraint test}, also known
as the Poisson equation. It is a key relation which we will use and test in this work. It writes as,
\be\label{eq:nondyn_con}
k^2\Phi=4\pi G a^2\rho_m\delta_m\,.
\ee
Using previously defined relations, in principle both sides of
Eq.~\ref{eq:nondyn_con} are calibrated and can be readily used to test it.

\begin{figure}[t]
  \begin{center}
  \epsfysize=0.38\textwidth
  \epsfxsize=0.49\textwidth
  \epsffile{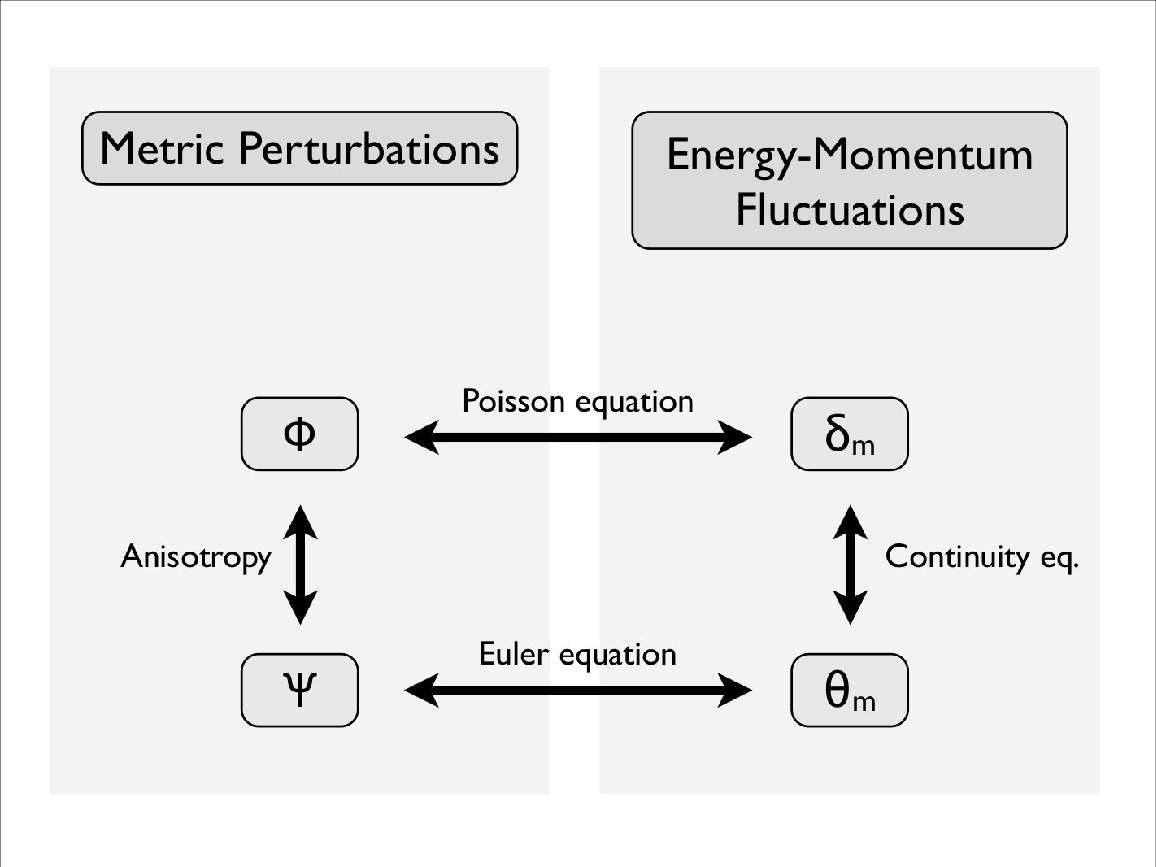}
  \caption{Web of cosmological tests of GR (see an analogous plot in
    Fig. 1 of \cite{Uzan:2006mf}).\label{fig:web}}
  \end{center}
\end{figure}

This set of relation constitute a web of possible cosmological as illustrated in Fig.~\ref{fig:web}.

As we hinted at before, to test any of this relations, we need various
observables. To trace matter perturbations, $\delta_m$, we will use biased tracers
like galaxies or clusters, whose relevant observational caveats will
be addressed below. Velocity surveys, direct (kinetic Sunyaev-Zel'dovich or peculiar velocities measurements) or indirect
(galaxy redshift survey) allow us in principle to probe
$\Theta_m$. Finally, metric perturbations are directly probed by weak
gravitational lensing which distorts the shape of source galaxies
along the line of sight. Since the geodesics are determined by the
gradient of combination of both $\Phi$ and $\Psi$, as $\nabla
(\Phi-\Psi)$, weak lensing will probe the integrated effect of metric
combination which is $2\nabla \Phi$. We will introduce our notations
for those observables below before discussing in more details the
specifics of the test we are interested in.

%--------------------
%--------------------

\section{Tracing matter perturbations with galaxies}
%===================================================
\label{sec:tracing_matter}

%If we are assured of systematic uncertainty of experiments, then it 
%would be general relativity test. If we are assured of general
%relativity prior, then it would be systematic uncertainty of
%experiments. Since weak lensing is supposed to be risky experiment due
%to the possible unknown source of systematic uncertainty, it can be
%regarded as systematic uncertainty test of weak lensing.

Whereas there is no direct probe of matter density fluctuations
(even though peculiar velocity measurements are usually considered to
be a direct tracer of density fluctuations, this requires the energy
momentum conservation law written in Eq.~\ref{eq:en_con} which is not
granted in our context), galaxies can still be used for this purpose
provided several observational artifacts are properly accounted for. 

Since we are working on large linear scales, we will ignore non-linear effects
which usually are another limiter. We will nevertheless discuss in the
next sections redshift distortions, redshift measurement uncertainties
and bias. The information extracted from galaxy density fluctuations
will be limited by how well we can handle those effects.  Because the
tests we are presenting require the use of projected quantities, we
will focus in particular on how those artifacts affect the projected
galaxy distribution within a redshift bin in its comparison to the
projected matter distribution. 

In this paper we will restrict ourselves to compare the measured and
predicted power spectra, even though a map-based approach would be
feasible. As we will work mostly on large scales where the field
distribution is Gaussian, considering this approach is lose-less. It is certainly not true on smaller scales.

\subsection{Cross-Power spectra conventions}
%-------------------------------------------

In this work, we are interested in cross-correlating various tracers
of the curvature perturbations at various redshifts to test
gravity. We will consider three observables based on gravitational
lensing, galaxy counts and velocity measurements from redshift-space
distortion. In particular, we will sometimes combine some projection
of those observables and we will thus define the associated angular
power spectra below.

We will consider that the Limber approximation is valid for all those
observables in the angular range of interest. The projected angular
power spectra $C^{XX'}_\ell$ of any pair of perturbations $X$ and $X'$ is
given by  
\ba\label{eq:cldd}
C_{\ell}^{XX'} & = & \frac{9}{25}\frac{2\pi^2}{(\ell+1/2)^3}\nonumber\\
& \times &\int dDD W^X(D) W^{X'}(D)\Delta_{\zeta\zeta}(a_0,k={\ell\over D})
\ea
where $\Delta_{\zeta\zeta}(a_0,k)$ is 
the rms amplitude of curvature fluctuations
on comoving hyper-surfaces at the starting epoch 
$a_0$ during matter domination and $D$ is the angular diameter
distance. The window function $W^X(D)$ is determined by the property of the quantity  $X$ as is shown explicitly below. 

We consider that galaxies follow a continuous distribution $n(z)$, defined as
\ba
n_g(z)\propto z^2 e^{-(z/1.5)^2}\,.
\ea
where we assume the space-based surveys~\cite{Bernstein:2003es}.
We consider that for each observed galaxy, we can measure either its spectroscopic redshift or its photometric redshift. We neglect the
errors on the former and we will discuss the latter further in Sec.~\ref{sec:bias_photo}. 
%assume for the latter Gaussian error with rms
%\be
%\label{eq:sig}
%\sigma(z)=\sigma_{\rm phz}(1+z)\ .
%\ee
%Then a subset $i$ of galaxies whose photometric redshift is such that $z_{i-1}<$$z$$<z_{i}$, follows the distribution \cite{Ma:2005rc}
%\ba
%n_i(z)={A_{i}\over 2}n_g(z)
%\left[ {\rm erfc}\left(\frac{z_{i-1}-z}{\sqrt{2}\sigma(z)}\right)
%-{\rm erfc}\left(\frac{z_{i}-z}{\sqrt{2}\sigma(z)}\right)\right]\,,\nn
%\ea
%where erfc is the complementary error function and $A_{i}$ is determined
%by a normalisation constraint. Throughout this work, we will choose
%$A_i$ so that $\int dz n_i(z) = 1$.

The galaxy over-density, $\delta_g$, are measured on the redshift slice labeled by $i$
at the comoving distance $D_i$ from the observer and is assumed to be
a linearly biased tracer of the matter over-density $\delta_m$, with
bias $b$, so that the window function for $\delta_{i,g}$ is given by
\ba
W^{i\,g}(D_i)=\frac{2}{3}G_{\delta_m}(a_i,k)\frac{dz_i}{dD}n_g(z_i)b(z_i)\frac{(l+1/2)^2}{\Omega_mH_0^2D_i^2}
\label{Eq:wig}
\ea
where the growth function $G_{\delta_g}$ is given by $G_{\delta_g}(a_i,k)=a
\Phi(a_i,k)/\Phi(a_0,k)$.  

The deflection angle ${\bf d}$ due to gravitational lensing is defined by 
the gradient field of the lensing potential, ${\bf d}={\bf \nabla} \phi$, where
\ba
\phi(z)=-2\int dD \int_z^\infty dz'
n_i(z')\frac{D(z')-D(z)}{D(z)}\phi(a,k)\ .
\ea
%where $D_i$ denotes the angular diameter distance to the source
%galaxies distributed on the thin redshift shell labeled by $s$.
The window function for $\phi$ is thus
\ba
\label{eq:winWL}
W^{i\,\phi}(D(z))=-2G_{\Phi}(a,k)\int_z^\infty dz'
n_i(z')\frac{D(z')-D(z)}{D(z')} ,
\ea
where the growth function $G_{\Phi}(a,k)$ is given by 
\ba
G_{\Phi}(a,k)=\Phi(a,k)/\Phi(a_0,k)\ .
\label{Eq:ga}
\ea
The angular power spectra of the deflection angle is given by
$C_\ell^{i\,dd}=l(l+1)C_\ell^{i\,\phi\phi}$. 

Finally, peculiar velocity in our context will be measured following the method proposed by \citet{Song:2008qt}.
On large angular scales, the evolution of LSS measured by peculiar velocity is an independent
tracer of the history of LSS. If we consider the energy momentum
conservation law written in Eq.~\ref{eq:en_con} to be valid, then the window function for
$\theta_{i,m}$ defined in the $i^{th}$ shell is given by 
\ba
W^{i\,\theta_m}(D_i)=\frac{2}{3}G_{\theta_m}(a_i,k)\frac{dz_i}{dD}n_g(z_i)\frac{(l+1/2)^2}{\Omega_mH_0^2D_i^2}\,,
\ea
where $G_{\theta_m} = a \dot G_{\delta_m}(a_i,k)$ is the growth factor
for $\theta^{}_m$ and the velocity bias is ignored,
$\theta_{i,g}=\theta_{i,m}$. 

\subsection{Redshift distortion and angular power spectra}
%---------------------------------------------------------

In this section, we consider a spectroscopic survey whose redshift
measurement errors are neglected and discuss the effects of redshift
space distortion on the angular power spectra. The galaxy density
fluctuations measured in redshift space are distorted by peculiar
velocities on all scales. The observed power spectra can thus be a
mixture of density fluctuations and peculiar velocities.  To derive
the projected galaxy angular power spectra on large scales,
i.e. linear, we first write the galaxy density in redshift space as
\cite{Kaiser:1987qv} 
\be
\delta_{g}(k,\mu,D) = \Phi(k,a_i)W_g(D)(1+f\mu^2)\ ,
\ee
where curvature perturbations are separated into a scale and a time
dependant part,
\ba
\Phi(k,a) = \Phi_o(k)\frac{\Phi(k,a)}{\Phi(k,a_o)}=\Phi_o(k) a G_{\delta_m}(a)\,,
\ea
and where $a_0$ is initial epoch at matter domination and $G_{\delta_m}$
is growth function of matter fluctuations and the window function
$W_g$ is defined in Eq.~\ref{Eq:wig}.
%The window function $W_g(D)$ is given by
%\be
%W_g(D)=\frac{2}{3}G_{\delta_m}(a)\frac{dz}{dD}n_g(z)b(z)\frac{(l+1/2)^2}{\Omega_mH_0^2D^2}
%\ee
%where $n_g$ and $b$ is number density of galaxy and galaxy bias respectively.
The fractional weight function between density fluctuations and peculiar
velocities $f$ can be written as
\be
f=\frac{d\ln G_{\delta_m}(a)}{d\ln a}\,.
\ee

The projected 2D galaxy density from redshift space is then written as 
\ba
\delta_{g}(\hatn) & = & \int dD W_g(D)(1+f\mu^2)\delta_{\Phi}^{}(\hatn,D)\\
& = & \int dD  \int {d^3\veck \over
  (2\pi)^3}W_g(D)(1+f\mu^2)\Phi(\veck,D)e^{ik\mu D}\, .\nn
\ea
We can then write
\ba
a_{\ell m}^g & = & \int d\Omega(\hatn) \delta_{g}(\hatn)Y_{\ell m}^*(\hatn)
\ea
hence
\begin{widetext}
\ba
(2\ell+1)C_{\ell}^{gg}&=& 
\sum_m \langle a_{\ell  m}^ga^{g*}_{\ell  m}\rangle\\
& = &  \sum_m\int d\Omega d\Omega'dD
dD'  {d^3\veck\over (2\pi)^3} {d^3\veck'\over  (2\pi)^3}\langle\Phi(\veck)\Phi^*(\veck') \rangle 
W_g(1+f\mu^2)W_g'(1+f\mu'^2)e^{i(k\mu D-k\mu'D')}Y_{\ell m}^*(\hatn)Y_{\ell m}(\hatn')\nn\nonumber\\
& =& \sum_m\int{d^3\veck\over (2\pi)^3}P_{\Phi}(k)\left[\int d\Omega dD W_g(1+f\mu^2) e^{ik\mu D}Y_{\ell  m}^*(\hatn)\right] \left[\int d\Omega'dD' W_g'(1+f\mu'^2) e^{ik\mu'D'}Y_{\ell m}(\hatn')\right]\ .\nonumber\\
\ea
\end{widetext}
We can then calculate the bracket as
\ba
\left[\ldots\right] &=& \int d\Omega dD W_g(1+f\mu^2)e^{ik\mu D}Y_{\ell  m}^*(\hatn)\nn\\
&= & \sqrt{\pi(2l+1)}\delta_{m0}\int d\mu dD W_g(1+f\mu^2) e^{ik\mu D}P_{\ell}(\mu)\nn\\
&=&2\sqrt{\pi(2l+1)}\delta_{m0}\int dD W_g(D)\nn\\
&\times&\left[j_l(kD)-f\left(j_{l-2}(kD)-\frac{l}{kD}j_{l-1}(kD)\right)\right]\ .
\ea
We thus obtain
\ba
C_{\ell}^{gg} = 4\pi\int{d^3\veck\over (2\pi)^3}P_{\Phi}(k)
\left[I_l(k)\right]^2
\label{eq:cl_gg}
\ea
where
\ba
I_l(k)&=&\int dD W_g(D) \\
&\times&\left[j_l(kD)-f\left(j_{l-2}(kD)-\frac{l}{kD}j_{l-1}(kD)\right)\right]\nn.
\label{eq:il}
\ea
At high $\ell$ modes where the Limber approximation is safe, we have
$kD\sim \ell$ which suppresses the contribution from the
$\Theta\Theta$ modes and we have
\ba
I_l(k)\simeq \int dD W_g(D) j_l(kD)\,,
\label{eq:cl_gg_limber}
\ea
where the redshift distortion effect on the projected power spectra is
suppressed. 
%Note again, this approximations is valid only on linear scales.

%--------------------
\begin{figure}[t]
  \begin{center}
  \epsfysize=0.49\textwidth
  \epsfxsize=0.49\textwidth
  \epsffile{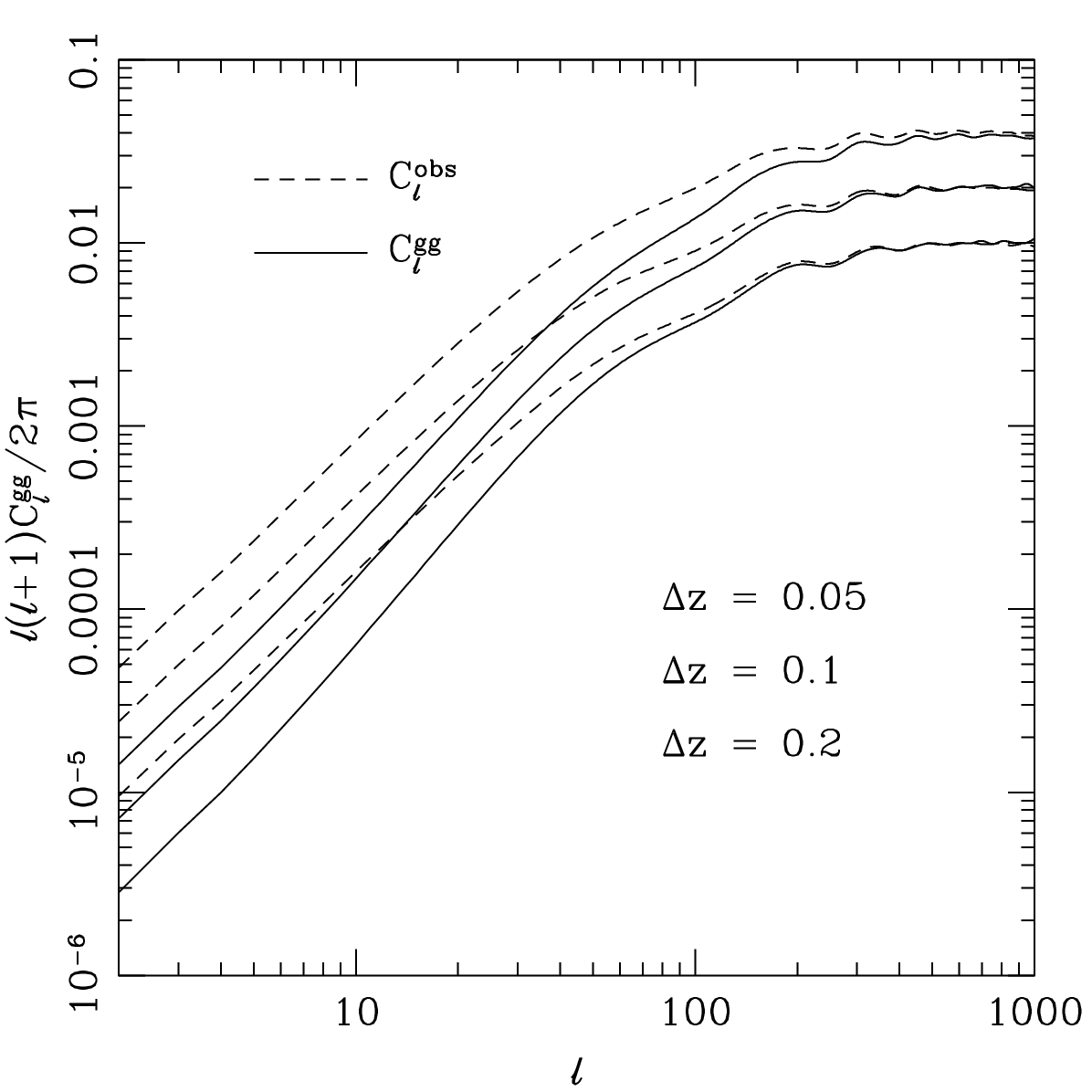}
  \caption{Projected galaxy over-density angular power spectrum as a function of multipole
    $\ell$ as defined in Eq.~\ref{eq:cl_gg} and Eq.~\ref{eq:il} (solid line)
    and its limber approximation as defined in
    Eq.~\ref{eq:cl_gg_limber} (dashed line). The various curves
    correspond to various bin width. Obviously, the wider the redshift
    bin the better is the Limber approximation but it is not a very
    accurate one except at smaller scales.
   \label{fig:cl_limber}}
  \end{center}
\end{figure} 
%--------------------

Those two equations illustrated in Fig.~\ref{fig:cl_limber} show that
projecting the angular power spectrum minimizes the redshift space
distortion through cancellation effects. This effect is
interesting here since it makes the projected angular density a more direct
tracer of the projected matter density. This will not be valid anymore
when non-linear effects are non-negligible, that is for scales larger than $k\sim 0.3 \,{\rm Mpc}^{-1}$. 

%--------------------
\begin{figure*}[t]
  \begin{center}
  \epsfysize=0.49\textwidth
  \epsfxsize=0.49\textwidth
  \epsffile{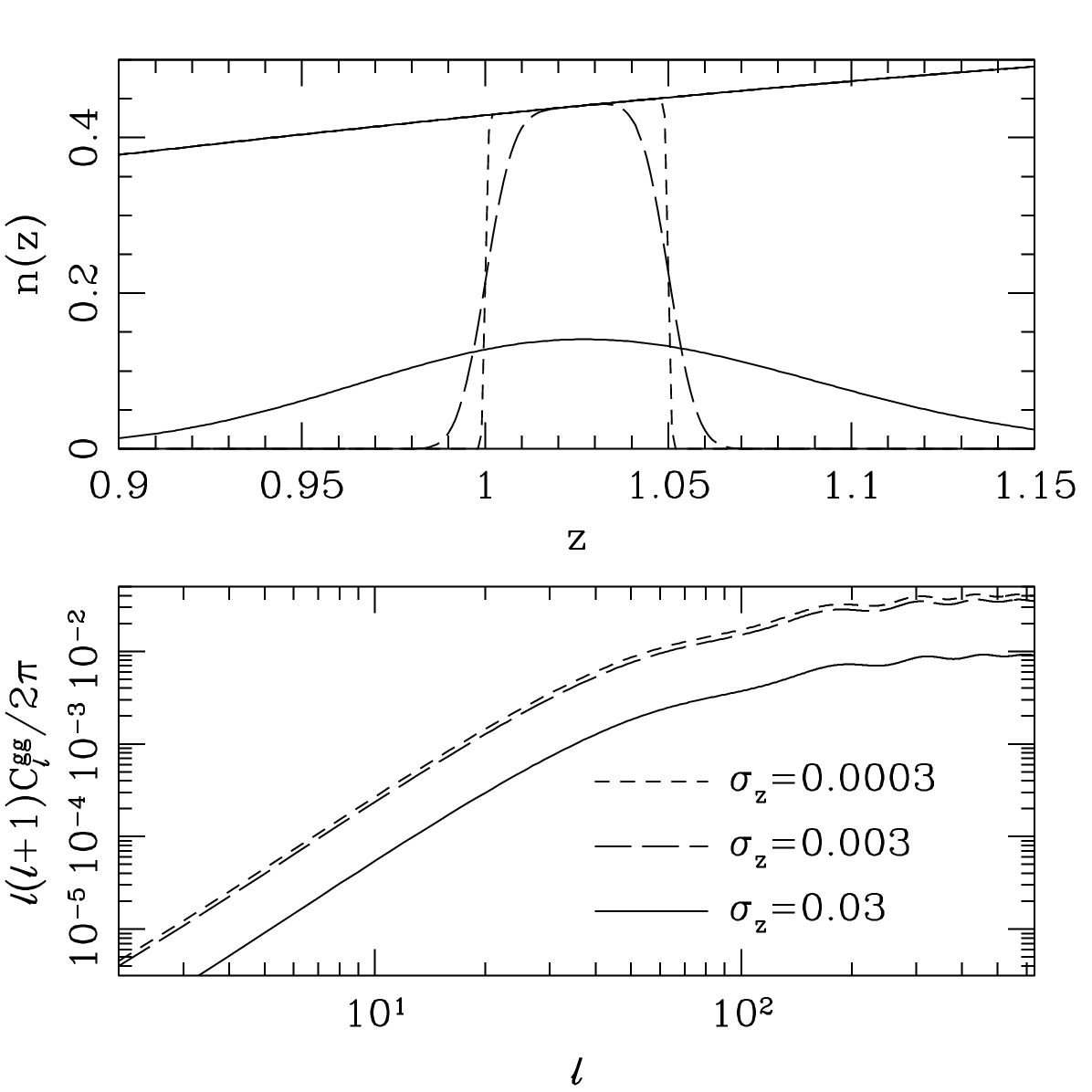}
  \epsfysize=0.49\textwidth
  \epsfxsize=0.49\textwidth
  \epsffile{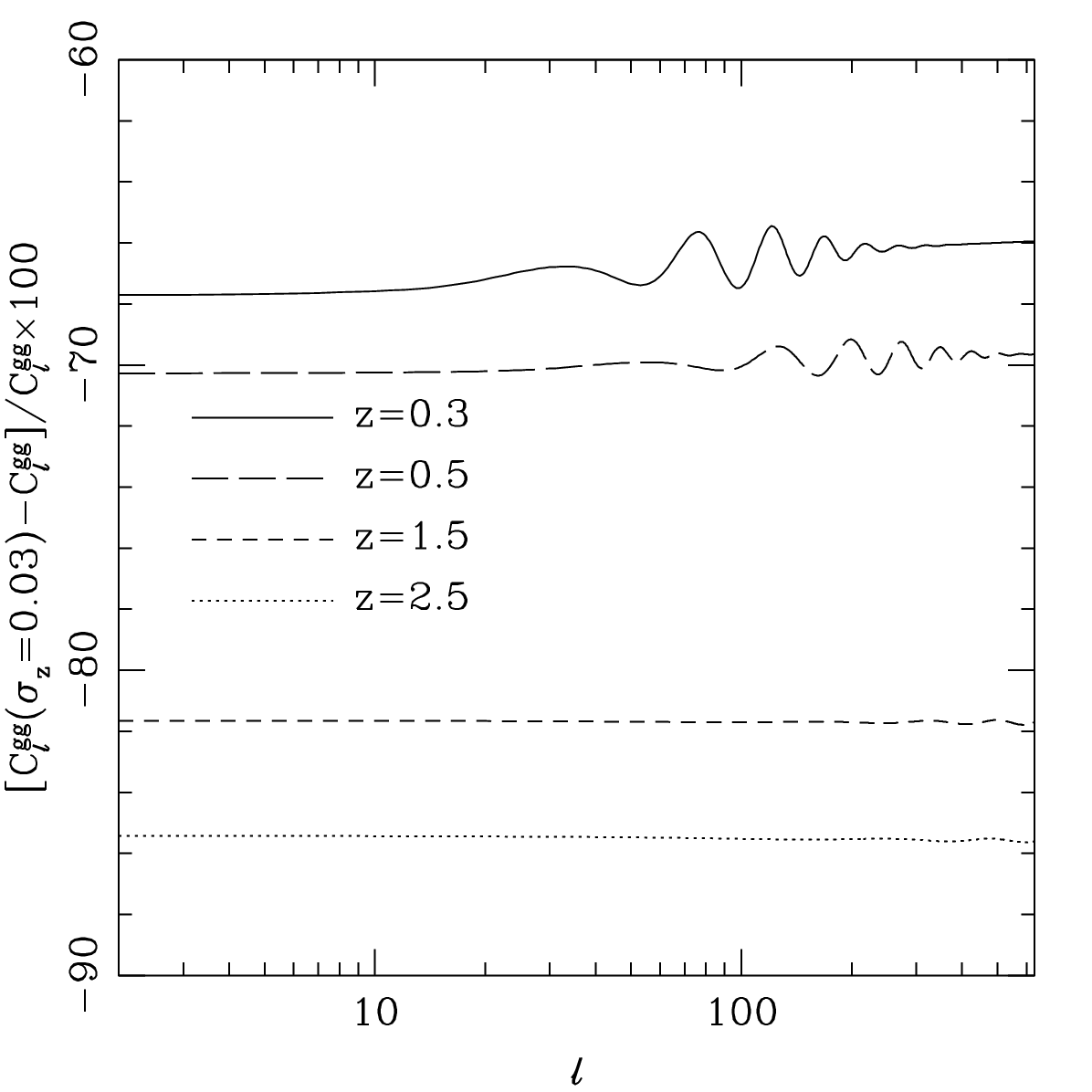}
  \caption{{\it Left panel, top:} $n_g$ distribution plotted for three
    various $\sigma_z$, \ie
    $3\times10^{-4},3\times10^{-3},3\times10^{-2}$ and
    the underlying galaxy distribution. {\it Bottom :} Corresponding
    galaxy overdensity angular power spectrum,
    $C_\ell^{gg}$. Photometric redshift error entails an important
    bias. {\it Right panel:} Relative difference between $C_l^{gg}$ power
    spectra with or without photo-$z$ errors considering
    $\sigma_z=0.03$. Solid, long dash, short dash and dotted curves
    correspond respectively to $C_l^{gg}$ at $z$=0.3, 0.5, 1.5 and
    2.5.}
   \label{fig:cl}
  \end{center}
\end{figure*} 
%--------------------

\subsection{Photometric uncertainties and angular power spectra}
%---------------------------------------------------------------

We now consider a photometric survey and discuss the effects of
redshift errors on the projected galaxy angular power spectrum. We
assume for the latter Gaussian error with rms 
\be
\label{eq:sig}
\sigma(z)=\sigma_{\rm phz}(1+z)\ .
\ee
A subset $i$ of galaxies whose photometric redshift is such that $z_{i-1}<$$z$$<z_{i}$, follows the distribution \cite{Ma:2005rc}
\ba\label{eq:wingal}
n_i(z)={A_{i}\over 2}n_g(z)
\left[ {\rm erfc}\left(\frac{z_{i-1}-z}{\sqrt{2}\sigma(z)}\right)
-{\rm erfc}\left(\frac{z_{i}-z}{\sqrt{2}\sigma(z)}\right)\right]\,,\nn
\ea
where erfc is the complementary error function and $A_{i}$ is determined
by a normalisation constraint. Throughout this work, we will choose
$A_i$ so that $\int dz n_i(z) = 1$.  Doing so, we discard the total number of galaxies as an observable.

As an illustration, we consider three different levels of photometric
errors, respectively $\sigma_z =$ $3\times10^{-4},$ $3\times10^{-3},$
$3\times10^{-2}$. While the first noise level seems idealistic, the
second one seems achievable in a near future
(e.g. \cite{Benitez:2008fs}) and the third one corresponds to what is currently achieved with SDSS \cite{Padmanabhan:2004ic}. The
corresponding window function $n_g(z)$ defined in Eq.~\ref{eq:wingal} is displayed in the left-top panel of
Fig.~\ref{fig:cl}. The left-bottom panel of Fig.~\ref{fig:cl} shows the simple projected
angular power spectra of the galaxy overdensity according to
Eq.~\ref{eq:cldd}. As we can see, the presence of photometric errors
introduces an important error that has to be taken into account. 

This error can be simply understood the following way. Looking at the
left panel of Fig.~\ref{fig:cl} one sees that whereas the mean of each distribution
and their integral -- the total number of galaxies in each bin -- are
identical, their variance are widely different. Since the curvature
perturbation power spectrum is weighted by the square of $n_g(z)$
spectrum when Eq.~\ref{eq:cldd} is applied to the galaxy overdensity, $\delta_g$,
this introduces the substantial bias we see in this plot. If the
true underlying distribution was known, i.e. if we could deconvolve
the observed distribution with the photometric error distribution,
then we could device easily weights that do not lead to such an
error. In practice though, it is unlikely that this deconvolution will
be feasible and we thus resort to another way to correct for this
effect by introducing another bias factor that we called a photo-$z$
bias, $b_z$.

Since $n_g$ is a smooth function of $z$ and since the matter power spectrum
is mostly featureless, we expect $b_z$ to be weakly dependent on scales.
In the right panel of Fig.~\ref{fig:cl}, we plot the relative
difference between the true galaxy angular power spectrum,
$C_\ell^{{\rm true}\,gg}$, and $C_\ell^{gg}$ with $\sigma_z=0.03$. 
In the low redshift bins, there are non-trivial scale dependent in the
angular range coming from baryonic features (100Mpc at $z=0.3$ corresponds roughly to $\ell=60$). In the
high redshift bins, the scale dependence is ignorable. Fortunately,
the contribution from those low redshift bins to the reconstructed
lensing power spectra is not significant. It means that we can treat
in a first approximation the biasing due to the limited photo-z error
as scale independent. Within this hypothesis, $C_\ell^{gg}$
is a linearly bias tracer of the matter angular power spectrum, with a
total bias $b_T^2=b_g^2b_z^2$, where $b_g$ denotes the bias due to
galaxy and $b_z$ denote the bias due to photo-z uncertainty. The
measure of the 3-D galaxy power spectrum allows us to measure $b_g$
separately but we can also measure directly the total bias in a
redshift bin, $i$, as will be detailed in Sec.~\ref{sec:bias_photo}.

\subsection{Galaxy bias and associated uncertainties}
%----------------------------------------------------

%-----------------
\begin{figure*}[t]
  \begin{center}
    \includegraphics[angle=0,width=0.49\textwidth]{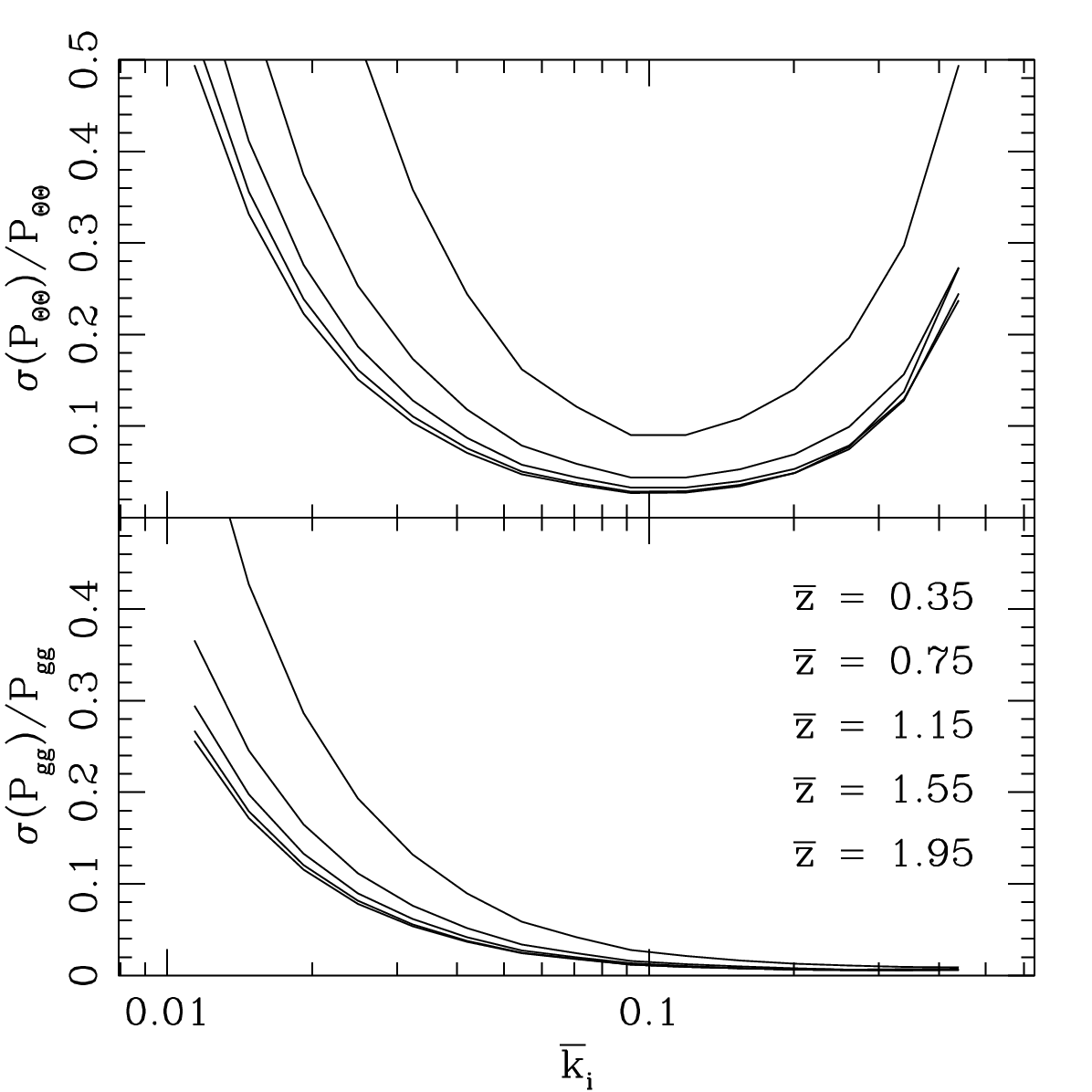}
    \includegraphics[angle=0,width=0.49\textwidth]{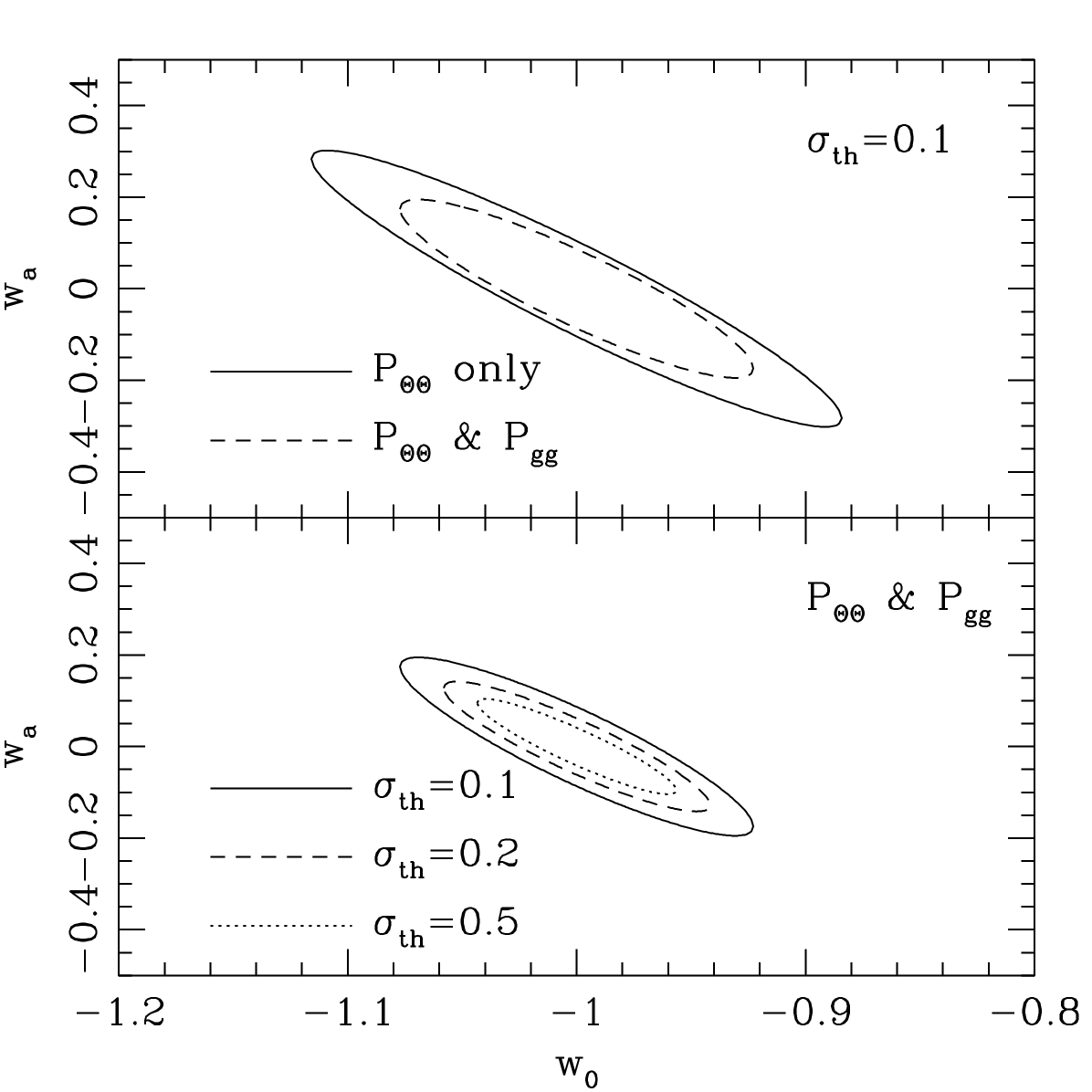}
    \caption{\footnotesize \emph{Left panels:} The top panel show the
      fractional errors for the reconstructed  $P_{\Theta_g\Theta_g}^{}$
      using the Fisher matrix formalism written in Eq.~\ref{eq:fish_gg_tt}. The
      bottom plot shows the corresponding fractional errors for $P_{gg}^{}$. \emph{Right panels:}
      68\% CL contour plots in the $w-w_a$ plane. The top panel
      shows the constraints obtained using $P_{\Theta_g\Theta_g}$ (no
      bias marginalization) only and
      $P_{\Theta_g\Theta_g}^{}$+$P_{gg}^{}$ (with bias
      marginalization). The bottom panel shows the constraints using
      $P_{\Theta_g\Theta_g}^{}$+$P_{gg}^{}$ and various value of the
      parameters $\sigma_{th}$ that quantifies the accuracy of the
      modeling of the Finger of God effect (see Eq. 15 of \citet{White:2008jy}
      for details).}
    \label{fig:contour_bias}
  \end{center}
\end{figure*} 
%-----------------

We now present two alternative ways to measure the linear galaxy bias,
assuming either a precise spectroscopic survey or a weak-lensing based
photometric survey.

\subsubsection{Using a spectroscopic survey}
%--------------------------------------------
\label{sec:bias_spectro}

Using precise spectroscopic redshift measurements, we are able to
separate the peculiar velocity power spectra from the
redshift-space power spectrum $P_g^{\rm obs}({\bf k})$ of a galaxy
redshift survey \cite{Song:2008qt}. The latter is commonly modeled as 
\ba
 P_g^{\rm obs}({\bf k}) &=& \left[ P_{gg}^{}({\bf k})
                    + 2\mu^2P_{g\Theta_g}^{}({\bf k})
                    + \mu^4P_{\Theta_g\Theta_g}^{}({\bf k})\right]\nn\\
 &\times&F\left(k^2\sigma_v^2(z)\mu^2\right) .
 \label{eq:pk_s1}
\ea
The separation of is  $P_{gg}({\bf k})$ and $P_{\Theta\Theta}({\bf
  k})$ made possible using the angular dependence of $P_g^{\rm
  obs}({\bf k})$ and where $\Theta=\theta/aH$ \cite{White:2008jy}.

We quantify how well this separation can be performed using a Fisher
matrix formalism. The Fisher matrix analysis relevant to this
separation for a given $k$ and $z$ bin is given by \cite{White:2008jy}
\ba
\lefteqn{F_{\alpha\beta}(k_i,z_j) = \int_{k_i^{\rm min}}^{k_i^{\rm
      max}}\frac{k^2dk}{2(2\pi)^2}\int^1_{-1} d\mu\nn\ V_{\rm eff}(k,\mu,z_j)}&&\\
&\times&\frac{\partial \ln P_{g}^{\rm obs}(k,\mu,z_j)}{\partial
  p_{\alpha}} \frac{\partial \ln P_{g}^{\rm obs}(k,\mu,z_j)}{\partial
  p_{\beta}}w_{FoG}^{}(k,\mu)\nonumber\\
\label{eq:fish_gg_tt}
\ea
where $\alpha$ and $\beta$ run from 1 to 2 and denote respectively $P_{gg}$ and
$P_{\Theta\Theta}$ and $w_{FoG}^{}$ is a down-weight function
filtering out the modes contaminated by non-linear redshift space distortions. Note that $P_{g\Theta_g}$ is
considered to be 1 in this separation which is valid on the linear
scales of interest to us. The effective volume $V_{\rm eff}^j$ in each
redshift bin $j$ is 
\be
V_{\rm eff}(k_i,\mu,z_j) = 
\left[\frac{n^jP_{g}^{\rm obs}(k_i,\mu,z_j)}{n^jP_{g}^{\rm obs}(k_i,\mu,z_j)+1}\right]^2
V_{\rm survey}(z^j)
\ee
where $n^j$ is the shot noise term coming from the finite galaxy
density supposed constant here, and $V_{\rm survey}(z_j)$ is the survey
volume in a given redshift bin. For the large scales of interest to
us, the cosmic variance term dominates over the shot noise and $V_{\rm
  eff}(k_i,\mu,z_j)$ is nearly identical to $V_{\rm survey}(z_j)$. For
our estimation, we will consider a full sky survey with a constant
galaxy density of $\bar{n}=5\times 10^{-3}\ h^3$Mpc$^{-3}$ and a constant
bias equals to 1. As illustrated in left panels of Fig.~\ref{fig:contour_bias}, we are able to
separate properly  $P_{gg}$ and $P_{\Theta\Theta}$ for wide $k$ and
$z$ bins.

Following this measurement of $P_{gg}$ and $P_{\Theta\Theta}$, we can
constrain simultaneously the cosmological parameters and the galaxy
biases. If the galaxy bias is scale-independent and depends only on
redshift, then the real space galaxy power spectrum $P_{gg}$ can also be written in terms of
the fundamental cosmological parameters plus a vector of bias parameters.
If we consider for example a survey up to $z=3.2$ with 8 redshift bins
of width $\Delta z=0.4$, the standard DE cosmological parameter set --
as the one used in the Dark Energy Task Force report
\cite{Albrecht:2006um} -- is extended to 16 elements, $q=(w,w_a,w_m,w_b,A_S,n_S,z_{\rm
  reion},\theta_S,b_{j=1-8})$. For this extended cosmological space,
the Fisher matrix simply writes
\ba\label{eq:fish_gt}
F_{mn}=\sum_{ij}\sum_{\alpha\beta}
\frac{\partial p_{\alpha}}{\partial q_m}
F_{\alpha\beta}(k_i,z_j)
\frac{\partial p_{\beta}}{\partial q_m}\ .
\ea
Since $P_{\Theta\Theta}$ is independent of bias, the degeneracy
between cosmological parameters and the bias is broken \cite{SongWhite08}, and we can
simultaneously measure in each redshift bin the bias and e.g. the dark
energy parameters. The resulting bias uncertainties are given for each
redshift bin in Table~\ref{table:bias} and are typically at the
percent level. Note that the galaxy bias is measurement is not
detrimental to the dark energy parameters as illustrated in the right
panels of Fig.~\ref{fig:contour_bias}.  

\begin{table}[thb]
\begin{center}
\caption{The fractional error of bias in some selected redshift bins using Eq.\ref{eq:fish_gt}.}
\begin{ruledtabular}
\begin{tabular}{c|ccccccc}
$z_j$ & 0.05 &  0.55  & 1.05  & 1.55  & 2.05 & 2.55 & 3.05  \\
\hline
${\Delta b_j\over b_j} (\%)$ & 0.75 & 0.37 & 0.33 & 0.38 & 0.45 & 0.53 & 0.57  \\
\end{tabular}
\end{ruledtabular}
\end{center}
\label{table:bias}
\end{table}

\subsubsection{Using a photometric survey}
%-----------------------------------------
\label{sec:bias_photo}

As introduced earlier, redshift uncertainties as big as the ones
resulting from photometric redshift measurements introduce an extra, almost linear bias, $b_z$. Since 
this bias does not affect the cross-correlation between galaxy
and peculiar velocity, we can  only determine the ``total bias'' by
cross-correlating the weak gravitational lensing signal and the
projected galaxy density. The signal-to-noise ratio for this correlation is simply given by~\cite{Acquaviva:2008qp}
\ba
\left(\frac{S}{N}\right)_j^2 & = &\sum_{s=1}^{8}\sum_{l=l_{\rm min}}^{l_{\rm max}} 
\frac{ f_{\rm sky}(2l+1) (C_\ell^{\delta_g^jd_s})^2}
{(C_\ell^{\delta_g^j\delta_g^j}+N_\ell^{\delta_g^j\delta_g^j})(C_\ell^{d_sd_s}+N_\ell^{d_sd_s})}\ ,
\nonumber
\\ 
&&
\label{eq:bz_snr}
\ea
and the fractional error on the galaxy bias is
\ba
\frac{\Delta b_t}{b_t} = \frac{1}{S/N}\ .
\ea
Some promising SNR numbers are given in Tab.~\ref{arr:bz_snr} where we used all
the $\ell$ up to $\ell = 500$, assumed $\sigma_z = 0.03$ and
considered a redshift binning for the WL signal from z=0 to 3.2 with
$\Delta z=0.4$. Note that for this evaluation, unlike in the previous
section we did not vary the other cosmological parameters but here
again, we expect the DE parameters to be non-degenerate with the bias
when we include the the projected galaxy and weak gravitational
lensing cross-correlation.

\begin{table}[htb]
\caption{Signal to noise ratio estimate for the total bias as defined
  in Eq.~\ref{eq:bz_snr} in selected bins.\label{arr:bz_snr}}
\begin{ruledtabular}
\begin{tabular}{c|c|c|c|c|c|c|c}
$z_j$ & 0.05 & 0.55 & 1.05 & 1.55 & 2.05 & 2.55 & 3.05 \\
\hline
$\left( S/N\right)_j$ & 160 & 430 & 300 & 170 & 88 & 35 & 6.6\\
$\Delta b_j/b_j (\%)$ &0.63 & 0.23 & 0.33 & 0.58 & 1.1 & 2.8 & 15 
\end{tabular}
\end{ruledtabular}
\end{table}

\section{Consistency tests}
%==========================
\label{sec:test}

\begin{figure}[t]
  \begin{center}
    %\epsfysize=3.0truein
    %\epsfxsize=3.0truein
    %\epsffile{cl_WL_WL_error.eps}
    %\epsffile{cl_WL_WL.eps}
    \includegraphics[angle=0,width=0.49\textwidth]{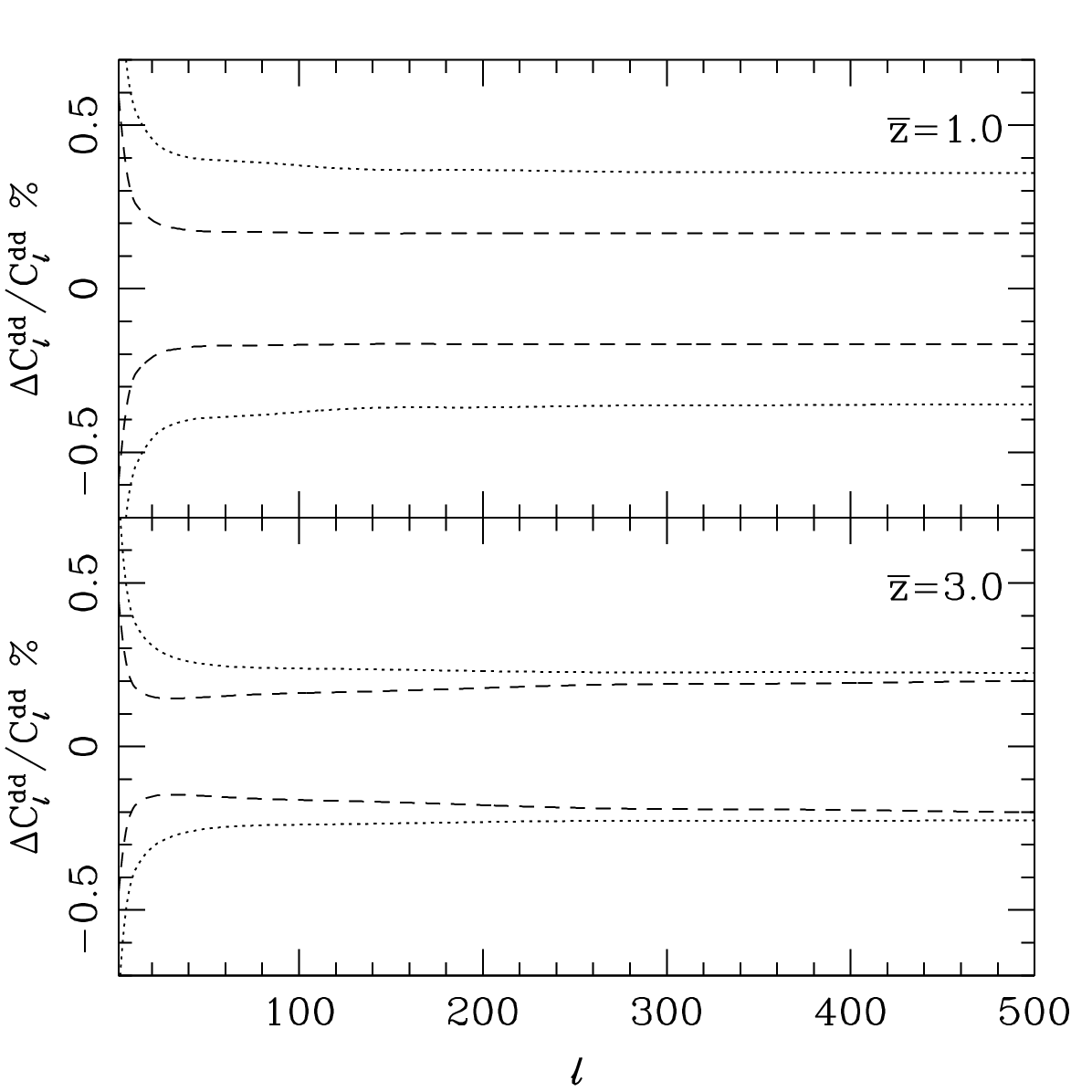}
    \caption{\footnotesize Relative uncertainties for the
      reconstructed $C_l^{dd}$. The dash line corresponds to the statistical uncertainties when measuring
      the bias using a large spectroscopic survey (as in Sec.~\ref{sec:bias_spectro})
      and the dotted line corresponds to the statistical uncertainties when using
      a photometric survey (as in Sec.~\ref{sec:bias_photo}). Other systematic bias are
      illustrated in Fig.~\ref{fig:cl_wl_wl_error}.}
    \label{fig:cl_wl_wl_rec}
  \end{center}
\end{figure} 

Now that we have presented how to obtain accurate estimates of the
projected matter angular power spectrum using galaxy surveys, we
proceed to the core of our study, that is the details of our cosmological
consistency tests. We will propose two tests. Either we predict the
lensing convergence power spectra using a galaxy survey and compare it to the
measured lensing power spectra, or we predict the cross-correlation
between matter and galaxy. While the first constitutes an
observational implementation of both the {\it metric test} written in
Eq.~\ref{eq:me_con} and the {\it non-dynamical constrain test}, the
second one is a direct implementation of the {\it non-dynamical constrain test} written in
Eq.~\ref{eq:nondyn_con}.

\subsection{Predicting the lensing power spectra}
%------------------------------------------------

\begin{figure*}[t]
  \begin{center}
    %\epsfysize=3.0truein
    %\epsfxsize=3.0truein
    %\epsffile{cl_WL_WL_error.eps}
    %\epsffile{cl_WL_WL.eps}
    \includegraphics[angle=0,width=0.49\textwidth]{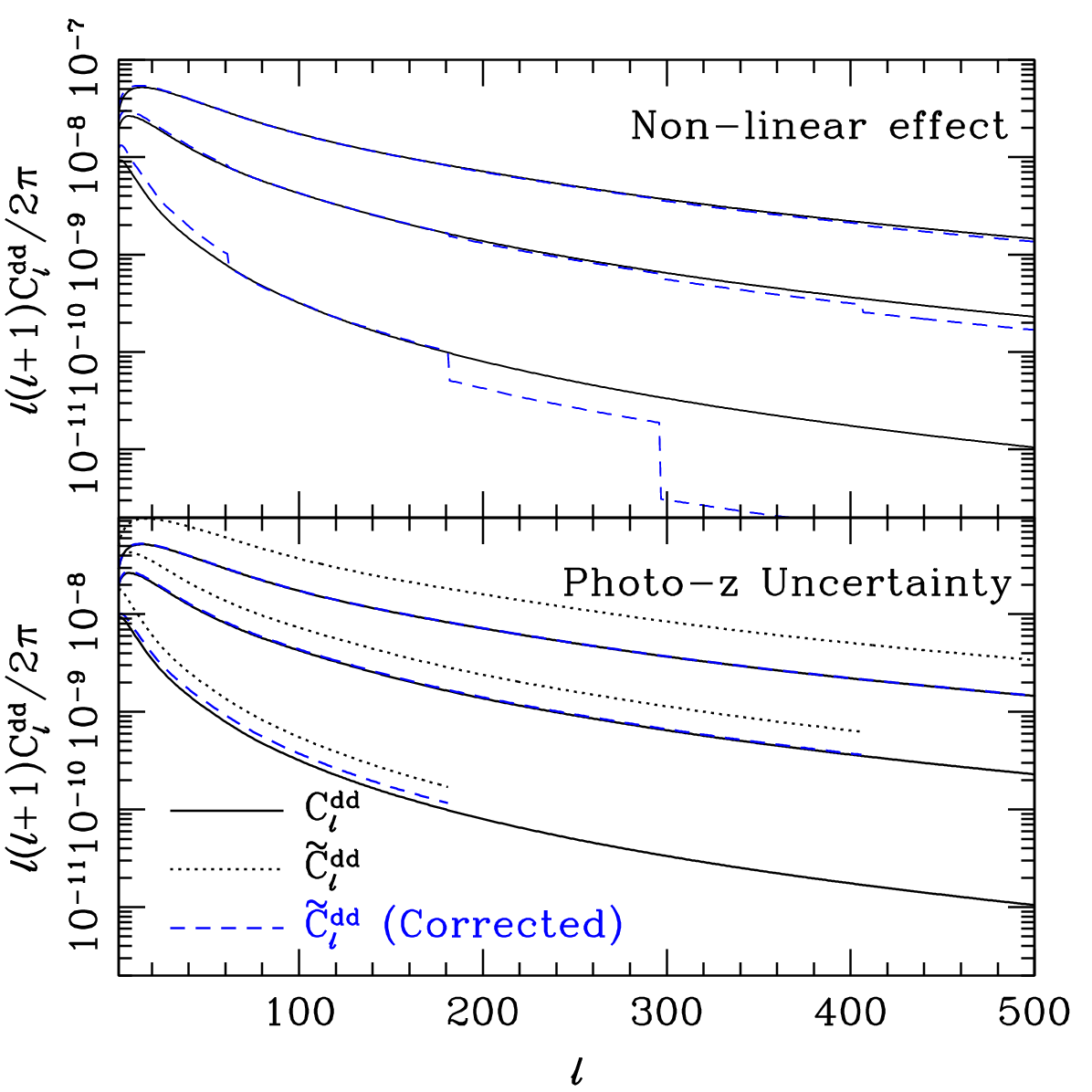}
    \includegraphics[angle=0,width=0.49\textwidth]{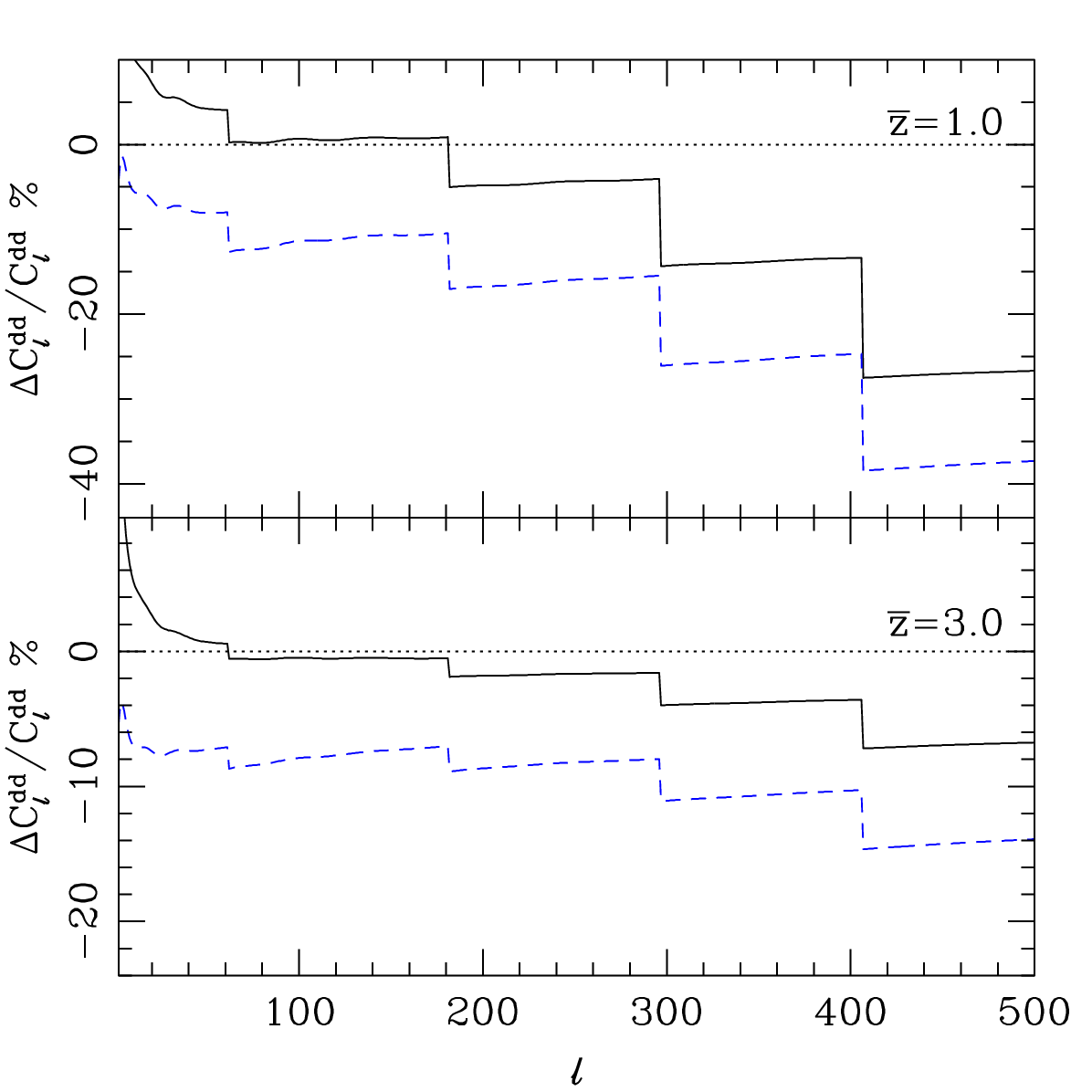}
    \caption{\footnotesize \emph{Left panel:} $C_l^{dd}$ for a source
      distribution at $\bar z$=0.2, 1.0 and 3.0 (from bottom to top)
      with a redshift width $\Delta z=0.4$. ({\it top panel}) The solid curves represent the
      original WL power spectra, and the dash curves represent the
      reconstructed one when the projected galaxy angular power
      spectra have been cut below some non-linear threshold chosen
      here to be $k\sim 0.3 Mpc^{-1}$.({\it bottom panel}) The solid
      curves represent the original WL power  spectra, the dotted
      curves represent the reconstructed ones with  photo-$z$ error of 
      $\sigma_z=0.03$ and the dash curves represent the predicted ones
      corrected for the photo-z bias. The agreement is obviously good
      except at low $z$ were the photo-$z$ bias is harder to correct 
      for. \emph{Right panel:} Expected relative uncertainties in the
      lensing predicted angular power spectra for the source
      distribution at two various redshifts (1.0 and 3.0 respectively from top to bottom
      panels and again $\Delta z=0.4$). The thin curves represent the
      reconstructed lensing signal using all the information available
      from the projected galaxy density. The visible noise at low
      $\ell$ originates in the finite number of galaxy redshift bins
      available. The thick curves is obtained when removing all the
      galaxy information above  $k\sim 0.3 Mpc^{-1}$ to stay in the
      linear regime. The black curves correspond to the reconstruction
      for a $\Lambda$CDM model. The blue curves to the reconstruction
      for $f(R)$ gravity with $B_0 = 1$ (see Sec.~\ref{sec:disc} for more
      details). Whereas the reconstruction performs really well in the
    }
    \label{fig:cl_wl_wl_error}
  \end{center}
\end{figure*} 

The density perturbations are measured on the redshift shell labeled by $i$
at the comoving distance $D_i$ from the observer. In the approximation
of a quasi-static evolution of perturbations, i.e. considering the
perturbations constant within a redshift shell, the projected angular
power spectrum can be written as 
\be
C_{l}^{i \; gg} = \frac{2 \pi^3}{(l+1/2)^3} 
\Delta D_i D_i W^{g}(D_i) W^{g}(D_i)\Delta_{\Phi\Phi}(a_i,k)\ .
\ee
Similarly, the weak lensing power spectra can be discretized as
\ba\label{eq:disccldd}
C_\ell^{s\,dd}
=\frac{2\pi^2}{l+1/2}\sum_{i=1}^{n}\Delta D_i D_i
\frac{4(D_s-D_i)^2}{D_s^2D_i^2}\Delta_{\Phi\Phi}(a_i,k)\ .
\ea

If we first assume that there is no dark energy perturbations, then
$\Delta_{\Phi\Phi}(a_i,k)$ can be written in terms of the angular
power spectrum of galaxies within this shell as 
\ba
\label{eq:phi_phi_}
\Delta_{\Phi\Phi}^{i}&=&\frac{9}{8\pi^2(l+1/2)}\frac{D_i^3}{\Delta D_i}
\left(\frac{dz}{dD}n_ib_i\right)^{-2}
\frac{\Omega_m^2H_0^4}{a_i^2}C_\ell^{i\,gg}.\nonumber\\
&&
\ea
Substituting this into Eq.~(\ref{eq:disccldd}), we are lead to define the reconstructed lensing power spectra
\ba
\label{recon_m}
\tilde{C}_{\ell}^{s\,dd} & = & \sum_{i=1}^{n}\frac{1}{b^{i\,2}_t}F^i_\ell\\
F^i_l & = & \frac{9}{(\ell+1/2)^2}\frac{D_i^2(D_s-D_i)^2}{D_s^2} 
\left(\frac{dz}{dD}n_i\right)^{-2}\frac{\Omega_m^2H_0^4}{a_i^2}C_\ell^{i\,gg}\ .\nonumber 
\ea
Note that this estimator is the simplest we can device and that we
considered the noise to be negligible. We also ignore correlations
within redshift bins, which is true if they are wide enough. If these
hypothesis are not full-filed, it is straightforward to generalize our
estimator to handle those effects in an optimal manner. In
Fig.~\ref{fig:cl_wl_wl_rec} we show the statistical errors for the
reconstructed power spectra using our nominal survey. Obviously, the
reconstruction performs very well. In the right panel of Fig.~\ref{fig:cl_wl_wl_error}, we plotted several 
reconstructed power spectra, before and after photo-z bias
reconstruction for $\Delta z=0.4$ bins and photo-z errors defined by
$\sigma_z= 0.03$. Obviously, the reconstructed estimator agrees well
with the input ones once corrected from the photo-z bias. As expected
following the results of Fig.~\ref{fig:cl} though, this bias is harder to
correct at low z. 

Once this estimator is defined, we can calculate the variance of $\Delta\tilde{C}_{l}^{s\,dd}$ as
\be
\Delta\tilde{C}_{l}^{s\,dd}=
\left\{\sum_{i=1}^{n}\left[\frac{1}{b_i^2} F^i_l \left(2\frac{\Delta b_i}{b_i}
\right)
\right]^2
\right\}^{1/2}
\label{Eq:cl_wl_wl_error}
\ee
which gives a fractional error 
\ba
\frac{\Delta\tilde{C}_{l\,I}^{s\,dd}}{\tilde{C}_{l\,I}^{s\,dd}}=
\frac{\left\{\sum_{i=1}^{n}\left[\frac{1}{b_i^2} F^i_l \left(2\frac{\Delta b_i}{b_i}
\right)
\right]^2
\right\}^{1/2}}{\sum_{i=1}^{n}\frac{1}{b_i^2} F^i_l}\ .
\label{Eq:rel_error_cl}
\ea

We show the predicted lensing signal accuracy in Fig.~\ref{fig:cl_wl_wl_error} using the same survey parameters as for
Fig.~\ref{fig:contour_bias}. The resulting statistical uncertainties
in the predicted angular power spectrum (right panel)  are much
smaller than the potential reconstruction biases from non-linear
effect, limited photometry measurement accuracy and discreteness
effects.  Percent accuracy is still possible with coming surveys and
we will discuss further in Sec.~\ref{sec:disc} the subsequent cosmological interpretation. 

The number of galaxy redshift bins used to approximate the continuous
kernel of lensing potential along the line of sight in
Eq.~\ref{recon_m} is limited. The choice of the optimal bin width is
motivated by several issues. We chose the thickness of the bins to be
larger than any correlation length between two subsequent
bins. Besides, since the projected galaxy angular power spectra are
defined in redshift space, a larger width smooth the redshift
distortion effect as shown in Fig.~\ref{fig:cl_limber}. We find a
width of roughly $\Delta z=0.1$ to be satisfying.  The thin curves in
the right panel of Fig.~\ref{fig:cl_wl_wl_error} shows that whereas the accuracy of
$\bar{C}_l^{dd}$ at $\bar z=0.2$ is limited by discreteness effect (we
can use only 4 z bins), it is nearly ignorable at high redshift
lensing bins. 

Another important bias factor involves non-linear effects. Whereas the
Poisson equation we used to predict the lensing signal is valid on all
scales, to reconstruct a reliable projected density template from the
galaxies measured in redshift space might be challenging in the
non-linear regime. To illustrate how important this effect is, we
filter out the galaxy template for $k<0.3h/Mpc$ and then project the
galaxy density field. In the right panel of
Fig.~\ref{fig:cl_wl_wl_error}, the stepwise curves represent the
resulting predicted lensing signal. Because non-linearities are
stronger at low redshift, this bias is more important when predicting
the lensing signal at lower redshift.

The uncertainty due to photometry measurement can be another
significant bias as shown in the left-bottom panel of
Fig.~\ref{fig:cl_wl_wl_error}. However, we presented in
Sec.~\ref{sec:bias_photo} how to deal with such an effect using
spectroscopic surveys or photometric surveys. This bias should thus
not the practical applications of our test.

\subsection{Adding cross-correlation with velocity}
%--------------------------------------------------
\label{sec:cc_with_vel}

\begin{figure}[t]
  \begin{center}
    %\epsfysize=3.0truein
    %\epsfxsize=3.0truein
    %\epsffile{cl_WL_WL_error.eps}
    %\epsffile{cl_WL_WL.eps}
    \includegraphics[angle=0,width=0.49\textwidth]{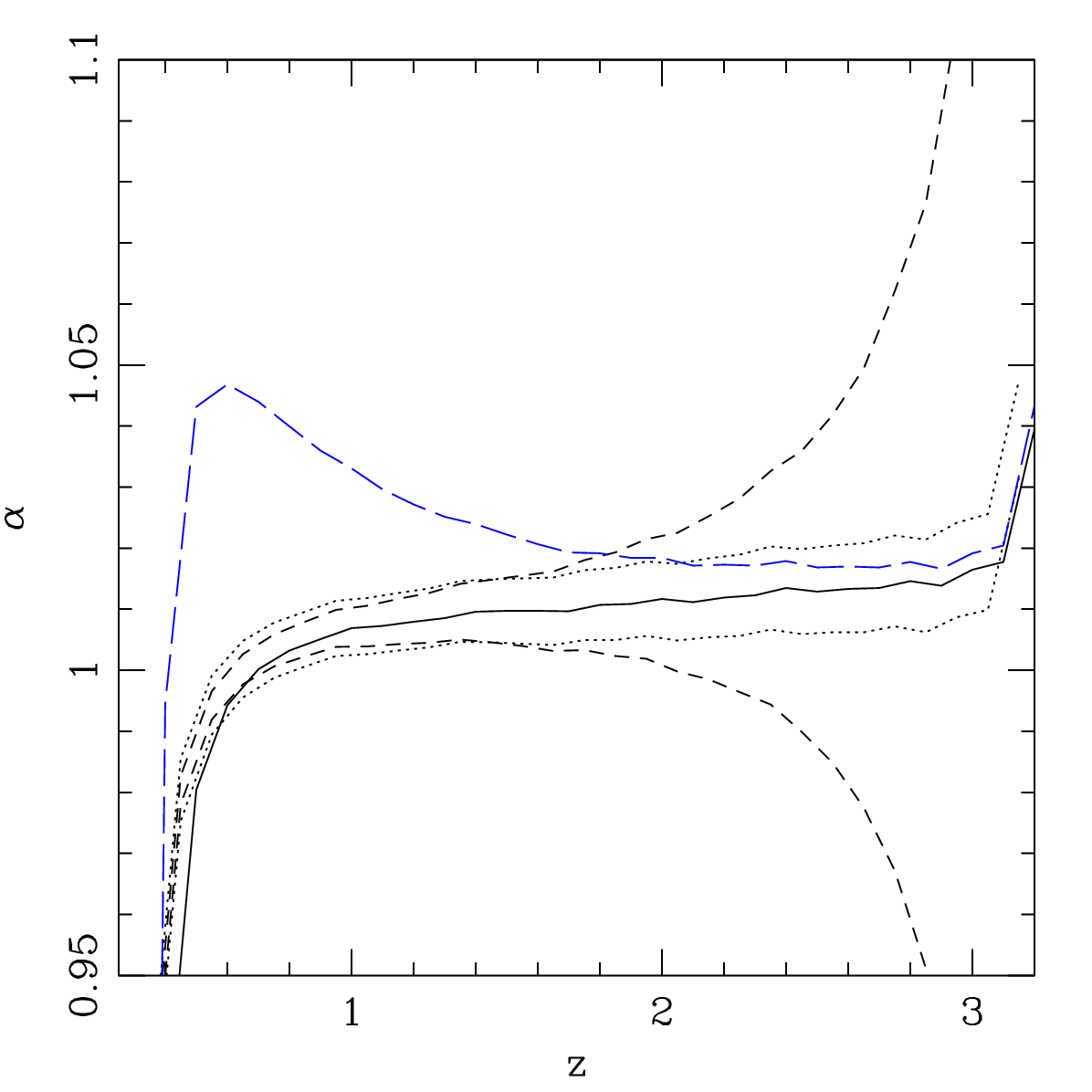}
    \caption{\footnotesize Black solid curve represent $\alpha_i$ for
      $\Lambda$CDM and blue solid curve represent $\alpha_i$ for f(R) theories
      with $B_0 = 1$. 
      Dash curves are errors estimated when using WL-galaxy correlations
      (as in Sec.~\ref{sec:bias_photo}) and dotted curves are
      error estimated using the galaxy-peculiar velocity correlations
      (as in Sec.~\ref{sec:bias_spectro}).} 
    \label{fig:alpha_error}
  \end{center}
\end{figure}

In addition to the comparison between the predicted and the measured
lensing power spectra at various redshifts, we can device another
consistency test using cross-correlations between galaxy and weak
lensing.  This test will highlight in particular  any deviation from
the Poisson equation that we parametrize with the $\alpha$ parameter defined by  
\ba
k^2\Phi= 4\pi G_N \alpha a^2 \rho_m \delta_m\ .
\ea
$\alpha = 1$ corresponds to GR. An estimator for $\alpha$ can be simply
derived using $C_\ell^{d\delta_g}$ and $C_\ell^{\delta_g\delta_g}$,
that we rewrite as
\ba
C_\ell^{d\delta_g} & = & \frac{2\pi^3}{l^2}\Delta D_iD_i\left(-2G_{\Phi}\frac{D_s-D_i}{D_sD_i} \right)\nonumber\\
&\times& \frac{2}{3}G_{\Phi}\frac{dz}{dD}n_ib_i\frac{al^2}{\alpha\Omega_mH_0^2D_i^2}\frac{9}{25}
\Delta_{\zeta\zeta}\\
C_\ell^{\delta_g\delta_g}& = & \frac{2\pi^3}{l^3}\Delta
D_iD_i\left(\frac{2}{3}G_{\Phi}\frac{dz}{dD}n_ib_i\frac{al^2}{\alpha\Omega_mH_0^2D_i^2}\right)^2\nonumber\frac{9}{25}\Delta_{\zeta\zeta}\ .\nonumber\\
\ea
The estimator $\alpha_i^{}$ is given by,
\ba
\alpha_i=-\frac{dz}{dD}n_ib_i\frac{l^2}{3\Omega_mH_0^2D_i}\frac{D_s}{D_s-D_i}
\frac{C_\ell^{d\delta_g}}{C_\ell^{\delta_g\delta_g}}\nn\ .
\ea
As such, the fractional error for $\alpha_i$ is equivalent to the
fractional error for $b_i$ given in Tab.~\ref{table:bias},
\be
\frac{\Delta\alpha}{\alpha}=\frac{\Delta b_i}{b_i}
\ee
and also we can also relate it directly to the compare it with the
fractional error of $\tilde{C}_{l}^{s\,dd}$ using
Eq.\ref{Eq:rel_error_cl}.

In Fig.~\ref{fig:alpha_error}, the estimated $\alpha$ from galaxy maps
and cross correlation with lensing potential is presented. Solid curve
is the estimated $\alpha$ which departs from unity around edges of
lensing kernel sourced by galaxies at $\bar z=3.0$. But the estimation
of $\alpha$ is very close to what is expected from GR model,
unity. Due to diverse bias effect, it is not identical to unity, but
it is just off by a percentage level. Still the difference due to
systematic effect is much smaller than the theoretical deviation due
to violated GR assumptions,  e.g $f(R)$ (R:Ricci scalar) gravity model with $B_0=1$
(long dash curve in Fig.~\ref{fig:alpha_error}),
where $B_0$ is defined as,
\be
B_0=\frac{d^2f/dR^2}{1+df/dR}\frac{dR}{d\ln a}\left(\frac{d\ln H}{d\ln
  a}\right)^{-1}\,.
\ee
Dash and dotted
curves are statistical errors from different galaxy bias estimation
from redshift survey alone and cross-correlations.
The advantage of this test over reconstructed $\bar{C}_l^{dd}$ is that
we are able to see tomographic view of consistency relation.

\section{Discussion}
%===================
\label{sec:disc}

In this paper, we advocate the use of consistency relations to test
gravity on cosmological scales. We did so by using a combination of
observables. This approach remains model independent and does not rely
on any specific parametrization. We focused in particular on the joint
use of galaxy surveys and weak-lensing observables. We showed how
using the former to predict the latter, we can build a strong
self-consistency test for GR. We also showed how 
large-scale redshift surveys can be extremely valuable in such an
endeavor. The test we proposed seems particularly appealing since any
weak-lensing survey is also by nature a galaxy clustering survey and
we illustrated how even a photo-z survey can be used to build a strong 
self-consistency test. We thus expect this test to be applied in the near
future to the CFHTLS survey \footnote{{\texttt
  http://www.cfht.hawaii.edu/Science/CFHLS/}} and others like DES
\footnote{{\texttt https://www.darkenergysurvey.org/}}, the proposed JDEM \footnote{{\texttt http://jdem.gsfc.nasa.gov/}} and
Euclid \footnote{{\texttt
    http://sci.esa.int/science-e/www/object/index.cfm?fobjectid=42266}}
space missions. 

We address in this work the key observational and astrophysical
systematics like the galaxy bias and redshift uncertainties but many
more survey specific ones would have to be carefully studied before any conclusion
can be drawn from any particular data-sets, e.g. $n(z)$ errors, PSF
correction and variation over large scales. It is quite likely that
this kind of test will be ultimately limited by the level of control
of systematics. In particular, we purposefully worked on linear
scales. Non-linear effects in real and redshift space will limit the
useful scales for such a program. If we assume that we can handle
non-linear effects
% -- for example the Finger of God redshift space
%distortion effect -- 
up to k=0.3 $h$/Mpc, we can see in the left panel
of Fig.~\ref{fig:contour_bias} that our program is still
tractable. Besides, weak-gravitational lensing has now been measured
in the linear regime where this test could be applied
\cite{Fu:2007qq}. Another unexplored sides of our work lie in the
possible degeneracies with underlying cosmological model parameters
but we leave this study for future work.

More positively, although, our survey parametrization is quite
generic, these surveys could certainly be customized to enhance the
discriminatory power of this test and others. We encourage future
missions to include the feasibility of such tests  as an extra criteria in the optimization of their design. 

Assuming that all systematics are well under control, we can wonder
what new physics can be probed with this test. As an illustration, we plotted in
Fig.~\ref{fig:cl_wl_wl_error} the predictions for a $f(R)$ theory where
a non-linear function $f$ of the Ricci scalar $R$ is added to the
Einstein-Hilbert action \cite{Song:2006ej}. We considered typical
value of the dimensionless Compton wavelength parameter $B_0=1$ (see Eq. 17 in
\cite{Song:2006ej}). In $f(R)$ gravity models, the Poisson
equation is modified so that $G_N \rightarrow G_N/(1+df/dR)$ and a non-zero anisotropy
stress is introduced to break the simplest GR model anisotropy
condition $\Phi=-\Psi$. As visible in Fig.~\ref{fig:cl_wl_wl_error}, the
predicted $\bar{C}_l^{dd}$ for a $f(R)$ gravity model using GR
assumptions would differ from the observed one at a level well exceeding the
various potential biases. This discrepancy originates from the GR
assumption in Eq.~\ref{recon_m} since any non-zero anisotropic stress
invalidates the relation $2\Phi=\Phi-\Psi$. However, the reduced
mass scale affecting $f(R)$ theories modify the Poisson equation. It
can thus be probed directly using the cross-correlation technique detailled in
Sec.~\ref{sec:cc_with_vel}. Note that since the difference between
this $f(R)$ model and  GR increases when the redshift decreases, so
does the discrepancy between the observed and predicted $C_l^{dd}$ for
a $f(R)$ theory. From the Fig.~\ref{fig:alpha_error} it can be
seen that constraints on $B_0$ of order $1$ or less are in principle
feasible with such a test. Other aspects of $f(R)$ theories are
discussed in \cite{Cooney:2008wk,Gannouji:2008wt,Nojiri:2008nt}.

%For reference, current upper limits on $B_0$ are of
%order 1 \cite{Song:2007da} so that a several order of magnitude
%improvement is foreseen using this test test. 

More generally, it can be argued that any theory with non-minimal interaction in the
dark sector or dark energy clumping would fail this test since it
either modifies the Poisson equation or introduce an anisotropic stress
\cite{Amendola:2007rr,Kunz:2006ca}. Reversely, it is important to note
that some types of modified gravity models would pass the test we
detailed here. For example, the potentials following from a
DGP model \cite{dvali00} are modified in such a way that the light
path tracing potential differences is similar to the standard
$\Lambda$CDM predictions. To detect such models, we need to
implement the program proposed in~\cite{Song:2008vm}. 

The tests presented above are only two of examples of the possible one
highlighted in Sec.~\ref{sec:conv} and we plan to extend our studies
to include other consistency relations as well as other observables
like the integrated Sachs-Wolfe effect \cite{Sachs:1967er} and the
lensing of the cosmic microwave background. We believe however that
the observational program sketched in this paper already offers a
stringent observational test of GR and we thus plan to put it into
action using current data.   

\acknowledgements{We thank Pat McDonald for useful discussions and
  Jean-Philippe Uzan for a careful reading of a draft version of this
  paper. YSS is supported by STFC.}

%\bibliography{modgrav}

\end{document}